\documentclass[12pt]{article}

\def\go{\;\&\;}

\def\ve{\vskip.5em}
\def\k{\kappa}

\def\vp{\4varphi}

\def\ora{\overrightarrow}

\def\Ra{\Rightarrow}

\def\half{\textstyle{\frac{1}{2}}}

\def\H{{\cal H}}

\def\vp{\varphi}

\def\H{{\cal H}}

\def\l{\lambda}

\def\ra{\rightarrow}
\def\tint{{\textstyle\int}}
\def\hg{{\hat g}}
\def\hp{{\hat\pi}}

\def\pa{\partial}
\def\dag{\dagger}

\def\b{\begin{eqnarray*}}  
\def\e{\end{eqnarray*}}    
\def\bn{\begin{eqnarray}}  
\def\en{\end{eqnarray}}   

\def\<{\langle}
\def\>{\rangle}

\def\no{\nonumber}

\def\ds{d^s\!x}
\def\k{\kappa}

\def\hk{\hat{\kappa}}
\def\{{\lbrace}
\def\hv{\hat{\varphi}}
\def\d3{d^3\!x}

\def\b{\beta}

\def\Oz{{\cal{O}}}

\def\}{\rbrace}

\begin{document}

\title{ How to Secure Valid Quantizations}
\author{John R. Klauder\footnote{klauder@ufl.edu} 
\\Department of Physics and Department of Mathematics  \\ 
University of Florida,   
Gainesville, FL 32611-8440}
\date{}
\let\frak\cal

\maketitle 
\begin{abstract} 
Classical mechanics involves position and momentum 
variables that must be  special coordinates chosen to promote to suitable quantum operators. Since classical variables may be broadly chosen, only  
unique variables should be chosen. We will outline how the favored variables and their suitable quantum operators is guaranteed to assure a truly valid quantization. Invalid quantizations may be  mistaken for valid ones, which then leads to incorrect physics. 
Besides particle examples,  there is also a brief 
run-through for fields and gravity.
\end{abstract}

\section{The Special Rules for a Valid \\Canonical Quantization}
In our search for  the proper classical variables, $p\go q$, and thus the proper variables to promote to quantum operators, $P\go Q
$, we introduce suitable coherent states. They are defined as
   \bn |p,q\>\equiv e^{-iq\,P/\hbar}\,e^{ip\,Q/\hbar}
   \,|\omega\> \;, \en
   where we choose the normalized fiducial vector, $|\omega\>$, so that $[Q+iP/\omega]\,|\omega\>=0$. 
   
   We first examine two important Hamiltonian functions given by 
   \bn && H(p,q)=\<p,q|\H(P,Q))|p,q\>=\<\omega|\H(P+p,Q+q)|\omega\>\no \\ &&\hskip3.4em
   =\H(p,q) +\Oz(\hbar; p,q) \:. \label{1} \en
 Although $p\times q$ has the {\it dimensions} of 
 $\hbar$ it follows that $|QP-PQ|=\hbar$ as well, and we may assume that $Q\go P$ can each be proportional to $\sqrt{\hbar}$. Adopting a general  polynomial style, with   $|\H(p,q)|<\infty$,  then the  term $\H(p,q$ has no $\hbar$, while the last term in (\ref{1}) contains all the factors with  $\hbar$. We now let $\hbar\ra0$, and  it follows  that $H(p,q)=\H(p,q)\Ra \H(P,Q)$, where we add what becomes the proper quantum Hamiltonian. That completes rule number one 
 for CQ.\footnote{In what follows, we often use CQ for canonical quantization, and we often use AQ for the second procedure, called affine quantization.}
 
 Rule number two, following Dirac \cite{dir}, will fix what kind of classical variables to choose for CQ. For that purpose, we again use the coherent states, as shown by
 \bn d\sigma^2_{CQ}=2\hbar [\:|\!|\;d|p,q\>|\!|^2-|\<p,q|\;d|p,q\>|^2\,]=\omega^{-1}\,dp^2+\omega\,dq^2\;, \en
 which requires a flat, Cartesian phase space of these two different variables, which may also be called a {\it constant zero curvature}. Observe, that this relation relies on $\hbar>0$, and is therefore a semi-classical relation. The second factor of  the coherent states is included to ensure that phase factors, such as $|p,q:f\>=e^{if(p,q)}\;|p,q\>$, do not 
 appear in the final metric. This safety  precaution is known as the  Fubini-Study metric,
  which was introduced in the early  19 hudreds \cite {fs}. 
  
  Clearly, using coherent states, 
   can help provide some tools to create a  valid quantization while using a canonical quantization.
   In using CQ, it is important to remember that $-\infty<P=P^\dag\go Q=Q^\dag<\infty$, which ensures   that the classical variables,  $-\infty<p\go q<\infty$, are also respected.
   
   \subsection{Finding the primitive ground state}
   Solving a simple ground state is elementary for canonical quantization, and we include this because our next quantization procedures will require a new formulation.
   
   Let us choose at first an elementary Hilbert space that only involves a finite coordinate space which is that of a circle of radius $R$. Net we introduce the operator $P$ and seek a wave function for which $P\;|
   \alpha\>=0$. Choosing Schr\"odinger's representation, the solution is $\<x|\alpha\>=c$, which is simply a constant. Now,  for a Hamiltonian that is simply $P^2\geq0$, it follows that $P^2\;c=0$, and since $\<\alpha|P^2|\alpha\> \geq0$, the `ground state' for this tiny problem is the constant $c$. We  can also choose  $c=1$, in which case $1=\Pi_y\,1$. Briefly stated, the ground state for the kinetic operator, $P^2$, is simply $1$.
   
   Of course, this state will change when we introduce a potential, $V(Q)$, and perhaps we can even expand the  radius $R\ra \infty$ to become the usual coordinate space.

Next, we introduce a newer quantization procedure, namely. AQ, that is a very useful tool for many different problems. The  next formulation of a `simple ground state' will then be something, very likely, quite new. 

   \section{The Special Rules for a Valid \\Affine Quantization}
   The fundamental difference between  AQ and   CQ is that, while $-\infty<p<\infty$ remains, we remove a coordinate point, say $q=0$, which then leaves $q<0$ and $q>0$. We  discard $q<0$ and keep only $q>0$. That implies that $Q=Q^\dag>0$, but that $P^\dag\neq P$, which can lead to complications. We can fix that by introducing the classical `dilation variable', $d=pq \Ra D=[P^\dag\,Q+Q\,P]/2=D^\dag$. Note that where $Q$ is sitting it forces any wave function  to vanish at the point removed; this property also permits $P^\dag\,Q=P\,Q$, which can be very useful for any calculations. 
   
 In addition, we choose to treat $q\go Q$ as dimensionless to simplify some equations. That means that both $p\go P$ and $d\go D$ have the dimensions of $\hbar$.
   
Comparing AQ and CQ, we 
find  that CQ has 
   $[Q,P]=i\hbar1\!\!1$,  while AQ has $[Q
   ,D]=i\hbar\,Q$. This latter equation resembles 
    the Lie algebra for the affine group,
  from which our procedures have adopted `affine' for its name.
  We are now prepared to introduce an AQ  paragraph
   to exhibit   the  differences from those of CQ.
   
   In our search for  the proper classical variables, $d\go q$, which lead  to the proper quantum variables, $D\go Q
$, we introduce suitable coherent states. They are defined as ({\bf Note:} $p,q\ra p;g$ in $|p:q\>$)
   \bn |p;q\>\equiv e^{ip\,Q/\hbar}\,e^{-i\,\ln(q)\,D/\hbar}
   \,|\beta\> \;, \en
   where we choose the normalized fiducial vector, $|\beta\>$, so that $[(Q-1\!\!1)+ iD/\beta\hbar]\,|\beta\>=0$, while $\beta>1/2$. Note that $-\infty<\ln(q)<\infty$, which involves the natural logarithm so that $e^{\ln(q)}=q>0$.
   
   We first examine two Hamiltonian-like expressions, denoted by a prime, e.g.,  $H'\go \H'$, given by 
   \bn && H'(pq,q)=\<p;q|\H'(D,Q)|p;q\>=\<\beta|\H'(D+pqQ,qQ) |\beta\>\no \\ &&\hskip4.1em
   =\H'(pq,q) +\Oz'(\hbar; pq,q) \;. \en
     In addition, 
we observe that any wave function that is even, i.e., $\psi(-x)=\psi(x)$, it follows that $\tint_{-\infty}^{\infty} (1-x)^p\,|\psi(x)|^2\;dx= 1+\Oz_p(\hbar)$, while the result is only $1$ after $\hbar\ra 0$ for a conventual classical Hamiltonian is $H=\half\,p^2 + V(|q|)$.

  Adopting a general  polynomial style, with   $|\H(pq,q)|<\infty$, and the final term $\Oz'(\hbar; pq,q) \ra0$ when $\hbar\ra 0$, it follows  that $H'(pq,q)=\H'(pq,q) \Ra\H'(D,Q)$, where again, the 
  quantum Hamiltonian is the correct choice. That is rule number one for AQ.
 
  Rule number two will fix what kind of classical variables to choose to promote to the valid AQ operators.
  For that purpose, we again use the coherent states, as shown here
 \bn d\sigma^2_{AQ}=2\hbar [\:|\!|\;d|p;q\>\,|\!|^2-|\<p;q|\;d|p;q\>|^2\,]=(\beta\hbar)^{-1}\,q^2\,dp^2+(\beta\hbar)\,q^{-2}\,dq^2\;, \en
 which has led us to  a {\it constant negative curvature} surface \cite{cnc}, a very special  phase space of  two variables.\footnote{The surface of a simple round ball acts as a {\it constant positive curvature}, which is related to spin operators, such as SO(3). With that, we have  fulfilled the three surfaces -- positive, zero, and negative -- while the negative  curvature cannot be seen like the other two. The spin coherent state stony is  covered in \cite{rrr}.}
 Observe, this metric relation also relies on $\hbar>0$, and is not a purely classical result. The second coherent state factor is included to ensure that no phase factor, such as $|p;q:g\>=e^{ig(p,q)}\;|p;q\>$, 
 appear in the metric. As stated before, this  safety  precaution  is known as the  Fubini-Study metric; see \cite{fs}.
  
  \subsection{Finding the new primitive ground state}
  For a system for which $q>0$ we have adopted affine variables for which the kinetic term becomes $p^2=d^2/q^2\Ra D(Q^{-2})D$. Now we seek the ground state for this operator. While $P\;1=0$, we find that $D\;x^{-1/2}=0$, again using Schr\"odinger's representation.
 It follows that an acceptable wave function, near $x=0$, is like $\gamma(x) =x^{3/2}\,(remainder)$ so that it is continuous and remains so after the first derivative. This particular expression will be explained further toward the end of the next section.
  
  Now, still using Schr\"odinger's representation, it follows that our ground state for  this kinetic operator is $\Pi_y\;y^{-1/2}$, and $D \;\Pi_y\;y^{-1/2}=0$, with an  eigenvalue of $0$, as before. As standard (positive) potential terms are now added, other eigenfunctions will emerge, and all others have larger eigenvalues. Now, a general wave function may be  written as $\gamma(x)=w(x)\,x^{-1/2}$, where $w(x)=x^2\,(remaider)$.
  Normalization leads to
  \bn  &&1=\tint\;|\gamma(x)|^2\;dx= \tint\; |w(x)|^2\;    \;dx/|x|\;.\en
  
  Clearly, relying on coherent states, 
   has provided fine tools to create valid quantizations when using CQ or AQ.

   \section{Two Toy Models: One for CQ, One for AQ}
  The two toy models are first a full-harmonic oscillator with classical Hamiltonian $H=(p^2 + q^2)/2$, while 
  $-\infty<p\go q<\infty$. The second toy model is the half-harmonic oscillator with the  same Hamiltonian, but while $-\infty<p<\infty$, now  $0<q$. Physically, in the second case, the particles bounce back off the `wall' at $q
 =0$ and  immediately changes direction, meaning that $p(t)$ changes sign immediately. That behavior is similar to someone throwing a tennis ball at a brick wall.
 
 An important item  of the harmonic oscillator quantum Hamiltonian, $\H =(P^2+Q^2)/2$, which, using CQ,  
  is the  eigenvalues , $E_n=\hbar(n+1/2)$, where $n=0,1,2,3,...$. Note that the 
  eigenvalues are {\it equally spaced by $\hbar$!}
 
 The second model, using AQ and $q>0$, first replaces the classical Hamiltonian with  $H'=(d^2/q^2+q^2)/2$ which then points to 
 \bn &&\H'=[D (Q^{-2})D+Q^2]/2 \no \\ \label{yy}
  &&\hskip1.5em =[P^2+(3/4)\hbar^2/Q^2+Q^2]/2 \:, \en  for which  the eigenvalues now have the form 
  $E'_n=2\hbar(n+1)$, with $n$ just  
  as  before. This implies that it is still {\it equally spaced}, now by $2\hbar$, with a ground state value of $2\hbar$ 
  \cite{lg}. Clearly, both of these two different quantizations are valid. 
  
  Although the operators $P^\dag\neq P$, because  $Q>0$,  the $\hbar$-term in (\ref{yy}) renders  {\it both} $P^\dag$ and $(P^\dag)^2$ to {\it act} like $P$ and $P^2$. Now every eigenfunction is of the form
  $\psi_n(x)=x^{3/2}(polynomial_n)e^{-x^2/2\hbar}$, as they were fully displayed in \cite{lg}.  To see what the second derivative does, take a break  and `solve' $[-\hbar^2\,(d^2/dx^2)+(3/4)\hbar^2/x^2]\;x^{3/2}=\,?$, for the interval $0<x<\infty$.
  
  Each toy model fully followed the rules for their  quantization procedures to achieve valid results. Interested readers might try to solve the  half-harmonic oscillator using CQ, and try to prove that it followed the rules; one example of that effort is presented in \cite{656}. 
 Otherwise, these particular results are most surely invalid, and will offer only incorrect physics.

  This concludes our simple examples with the fact that CQ can truly solve certain problems, while  
 AQ can truly solve other problems. Difficult problems in one procedure may find success in the other process. To build many valid examples just trade the potential $q^2/2 $ for a general potential 
 $V(q)\Ra V(Q)$, providing that it obeys $|V(q)|<\infty$.
 
 By just adding a different potential, you are certain to have valid results. The focus on getting things right is on the kinetic factor, i.e., $p^2\Ra ?$, and much less on the potential, $V(q)$.
 
 \section{The Benefits of an Affine Quantization  for Field Theories}
 Next, we study some common  classical Hamiltonians, such as 
 \bn H(\pi,\vp)=\tint\{\half[\pi(x)^2+(\ora{\nabla}\vp(x))^2+m^2\vp(x)^2]+g\,\vp(x)^p\}\;\ds\en 
 In a path integration of these models, there are cases where $|\pi(x)|$ and/or  $|\vp(y)|$ reach $\infty$, while still having $H(\pi,\vp)<\infty$.

  \subsection{A Brief Overview 
  of \\Quantum Field Theory}
 To begin with, we focus on a feature of mathematics that impacts physics. As one example, consider the function $f(x)=1/|x|^{1/3}$ in the interval  $-1<x<1$, where $f(0)=\infty$. It follows that $\tint^1_{-1} f(x)^2\;dx <\infty$, while $\tint_{-1}^1 f(x)^4\;dx=\infty$. Later, we will refer to this situation as an `f-issue'  which involves a field that reaches infinity, and also  is part of an integration that is finite.
 
 Physics is frequently engaged in studying nature, and then having any field with an f-issue, as in the last paragraph, seems impossible. Stated bluntly, the strength of any field of nature can not reach infinity. 
 The author accepts that no field of nature reaches infinity, and we now endeavors to make that happen in our analysis. For example,
 a classical Hamiltonian density, $H(x)$, that describes a part of nature, should not reach infinity for any $x$, and that fact needs to be part of the mathematics involved. 
 
 Presently, the mathematical focus only requires that $\int H(x)\;\ds<\infty$, and this approach can lead to a nonrenormalizable behavior if the interaction power of  terms, such as for $\vp^p_n$ in (9), when $p\geq 2n/(n-2)$. 
 Let us examine a procedure in which we can actually favor nature.
 
 \subsection{A typical model of a covariant scalar field}
 Our models of interest have the classical Hamiltonians, already introduced above, and once again  is
  \bn H(\pi,\vp)=\tint\{\half[\pi(x)^2+(\ora{\nabla}\vp(x))^2+m^2\vp(x)^2]+g\,\vp(x)^p\}\;\ds\;. \label{555}\en
  When such a model is quantized, say by a path integration procedure, the classical Hamiltonian is evaluated by a vast number of functions for both $\pi(x)$ and $\vp(x)$. In so doing, fields that can diverge but still offer finite integrations -- like the example of an f-issue in a previous paragraph -- are conventionally introduced along with fields without any divergencies. 
  
  How can we limit the classical fields so that f-issues do not arise? The answer to that question appears in the next section, and it is much easier than could have been expected.

  Following the simple rules of Sec.~2, we introduce the dilation field $\k(x)=\pi(x)\,\vp(x)$, along with $\vp(x)=0$, because if $\vp(x)=0$, then $\k(x)=0$ and $\pi(x)$ cannot help. But is it acceptable to remove $\vp(x)=0$?\ve

  {\bf Comment:} Consider, a rain storm that has  a rain quantity=(qu-r) per hour that is $0<(qu-r)<\infty$. If $(qu-r)=0$, there simply is no rain. If instead, it was a snow storm, a similar story could be  a snow quantity=(qu-s), where $0<(qu-s)<\infty$. In fact, the physics of both is identical when 
  $(qu-r)=(qu-s)=0$. So we ignore rain and snow when they are absent. 
  
  If $\vp(x)$ represents a physical particle, say an electron, the field can remove $\vp(x)=0$ if the particle is not present. Maybe math may be concerned, but physics 
 can certainly accept removing any portion of a field representation when that field  represents absolutely  nothing. \ve
  
  When using these  new classical variables, the classical Hamiltonian, from (\ref{555}), becomes
  \bn H'(\k,\vp)=\tint\{\half[\k(x)^2/\vp(x)^2 +(\ora{\nabla}\vp(x))^2+m^2\vp(x)^2]
  +g\,\vp(x)^p\}\:\ds \;,\label{666}\en
  Already we see that $0<|\vp(x)|<\infty$ for if  $\vp(x)^{2}=0$ or $\vp(x)^{-2}=0$, it means that $\k(x)$ fails to offer any value for  $\pi(x)$. It already follows that $|\k(x)|<\infty$ and  $0<|\vp(x)|^p<\infty$, and therefore, {\it nonrenormalizability vanishes!} 
  Such  field models clearly  
  keeps both sides of $\vp(x)\neq0$, i.e., both $\vp(x)>0$ and $\vp(x)<0$. Since there is the gradient term, the field will appear to be continuous and integrations should not be affected. 
  
  To ensure our quantization follows the rules, we introduce suitable coherent states \bn
  |\pi;\vp\>=e^{i\tint\,\pi(x)\,\hv(x)\,\ds/\hbar}\;e^{-i\tint\,\ln(|\vp(x)|)\,\hk(x)\,\ds/\hbar} \,|\beta\>\;, \en
  while  the new fiducial vector obeys  $[(|\hv(x)|-1\!\!1)+i\,\hk(x)/\beta]\,|\beta\>=0$, and  the same label now referees to a  new fiducial vector.
  
  It follows that
  \bn && H'(\k,\vp)=\<\pi;\vp|\H'(\hk,\hv)
  |\pi;\vp\> \no \\ &&\hskip4em
  =\<\beta|\H'(\hk+\pi|\vp|\,\hv, 
  |\vp|\,\hv) |\beta\> \no \\  && \hskip4em 
  =\<\beta|\H'(\hk+\pi\,\vp\,|\hv|,
  \vp\,|\hv|) |\beta\> \no \\ &&\hskip4em =
  \H'(\k,\vp)+\Oz(\hbar;\k,\vp)\:, \en
  which, as was like the case for particles, we find, if $\hbar\ra0$, then  $H''(\k,\vp)=\H''(\k,\vp)\Ra \H''(\hk,\hg)$.

  To offer an affine quantization for this example, we first introduce the dilation operator $\hk(x)=[\hp(x)^\dag \hv(x)+\hv(x)\hk(x)]/2$ and $\hv(x)\neq0$. Next, adopting a Schr\"odinger representation for the quantum Hamiltonian, we are led to
  \bn &&\H'(\hk,\vp)=\tint \{\half[\hk(x)\,(\vp(x)^{-2})\,\hk(x)+(\ora{\nabla}\vp(x))^2+m^2\vp(x)^2] \no\\
  &&\hskip14em +g\,\vp(x)^p\}\;\ds \:, \label{777}\en
  where it is evident that $\vp(x)^2>0$ and $1/\vp(x)^2>0$, and therefore $0<|\vp(x)|^p<\infty$, which eliminates nonrenormalizability, as desired. 
  Finally, we offer Schr\"odinger's equation as
    \bn i\hbar\,\pa\,\Psi(\vp,t)/\pa t = \H'(\hk,\vp)\,\Psi(\vp,t)\;.
    \label{888}\en
    
    As usual, it may be necessary to introduce some version of a regularization for these equations, but these same equations should point the way to proceed. To offer some support, we note that although   $\hp(x)^\dag \neq  \hp(x)$ it can be helpful to know that 
     $\hp(x)^\dag \vp(x)=\hp(x) \vp(x)$; see footnote 1.
    
    If the reader has already accepted the expressions for the half-harmonic oscillator in (\ref{yy}), they may be willing to accept (\ref{777}) and (\ref{888}) for this quantized covariant scalar model as well.\footnote{We note that Monte Carlo studies of the scalar fields $\vp^4_4$ and $\vp^{12}_3$ using canonical quantization have led to ``free-results'', as if the interaction term was absent when it was not. However, using affine quantization has led to ``non-free-results'', when the interaction term makes different results when  the coupling constant  changes; see [8 - 12 ].}

 \section{Applying Affine Quantization to \\Einstein's Gravity}
 In order to quantize gravity it is important to render a valid quantization of the ADM classical Hamiltonian \cite{adm}. We first choose our new classical variables that includes
 what we will call the dilation field $\pi^a_b(x)\equiv \pi^{ac}(x) \,g_{bc}(x)$ (summed by $c$) along with the metric field $g_{ab}(x)$. We don't need to impose any restriction on the metric  field because physics already requires that $ds(x)^2=g_{ab}(x)\;dx^a\;dx^b>0$ provided that $\{dx^a\}\neq0$; we choose that $g_{ab}(x)$  is  dimensionless because that will simplify further equations. 
 The metric field can also be diagonalized by non-physical, orthogonal matrices, and then it includes only $g_{11}(x), \;g_{22}(x), \;\& \;g_{33}(x)$, or what we will call   $g_{[ab]}(x)\equiv g_{ab\:   (diagonal)}(x)$ each element of which must be strictly positive as required by physics.\footnote{The reader should compare  the three diagonalized positive metric variables with $q>0$ in the half-harmonic oscillator story, which then required an affine quantization.}
 Now we introduce two diagonal matrices, $A\equiv g_{[ab]}(x)$ and  $B\equiv\eta_{[d]}^{[c]}(x)$.  Next we let $A=e^{B}$, and use $O$ and $O^T$, one field element from  $SO(3)$ and its transpose, and where $O^T(x)\,O(x)=O(x)\,O^T(x)=I$. Now, we are led to
 \bn \bar{A}\equiv O^TA\,O=O^Te^{B}\,O
 =e^{O^TB \,O}\equiv e^{\bar{B}} \;, \en
 where we now let $\bar{B}(x) =[\eta(x)]$ and
 $[e^{[\eta(x)]}]_{ab}=[g(x)]_{ab}\equiv g_{ab}(x)>0$.
 
 Moreover, we now have 
 \bn &&\hskip-.9em H''(\pi, g)=\<\pi;g|\H''(\hk,\hg)|\pi;g\> \no \\ &&\hskip3em =\<\alpha|\H''(\hk^a_b +\pi^{ac}g_{bc}\;\hg_{bc}, \,g_{m n} \; \hg_{mn})|\alpha\> \no \\ &&\hskip3em
 =\H''(\pi,g) + \Oz(\hbar;\pi,g)\;,\en
 where here  $\pi$ stands for $\{\pi^a_b(\cdot)\}$ and $g$ stands for  $\{g_{cd}(\cdot)\}$, and use the facts that $\<\alpha|\hg_{de}(x)|\alpha\>=\delta_{de}$ and $\<\alpha|\hp^c_d(x)|\alpha\>=0$. Since $\hp^a_b\propto \hbar$, it follows that if $\hbar\ra0$, then $H''(\pi, g)=\H''(\pi, q)\Ra \H''(\hp,\hg)$, pointing the way to a valid quantization of gravity.
 

 Next we present the ADM classical Hamiltonian  in our chosen affine variables, which, now introducing $g(x)\equiv \det[g_{ab}(x)]>0$, leads to
  \bn &&H''(\pi,g)=\tint \{g(x)^{-1/2}[\pi^a_b(x)\pi^b_a(x)-\half\pi^a_a(x)\pi^b_b(x)] \no \\
  &&\hskip10em +g(x)^{1/2}\,^{(3)}\!\!R(x)\}\;d^3\!x\;,\en
  where $^{(3)}\!\!R(x)$ is the Ricci 
  scalar for three spatial coordinates, and which contains all of the spatial derivatives of the metric field. Already this version of the classical Hamiltonian contains reasons that restrict $g(x)$ to
  $0<g(x)<\infty$, plus $|\pi^a_b(x)|<\infty$, and $|^{(3)}\!\!R(x)|<\infty$. Just like the field theory example using affine variables, our gravity fields have  no f-issues for the gravity story. 
  
  We next introduce the coherent states for gravity. For that we use the dilation gravity operator 
  $\hp^a_b(x)=[\hp^{ac}(x)^\dag\,\hg_{bc}(x)+\hg_{bc}(x)\, \hp^{ac}(x)]/2$ along with $\hg_{ab}(x)>0$. 
   Our gravity coherent states are
 \bn |\pi;g\>=e^{i\tint\,\pi^{ab}\,\hg_{ab}(x)\;\ds/\hbar}\;e^{-i\tint\; \eta^a_b(x)\,\hp^b_a(x)\;\ds/\hbar} \;|\alpha\>\;, \en
 where we choose $|\alpha\> $ by $[(\hg_{ab}(x) -\delta_{ab}1\!\!1)+i(\hp^c_a(x)+\hp^c_b(x))/2\alpha(x)\hbar]\;|\alpha\>=0$.
 
Using Schr\"odinger`s representation, we are led to
  \bn &&\H'(\hp,g)=\tint \{\;[\,\hp^a_b(x)\,g(x)^{-1/2}\,\hp^b_a(x)
  -\half \hp^a_a(x)\,g(x)^{-1/2}\,\hp^b_b(x)\;]\no \\
  &&\hskip10em +\,g(x)^{1/2}\;^{(3)}\!\!R(x)\;\}\;d^3\!x \:. \en
 Finally, we close with Schr\"odinger's equation
   \bn i\hbar\,\pa \:\Psi(g,t)/\pa t=\H'(\hp,g) \;\Psi(g,t)\:, \en
   which offers the necessary ingredients for the foundation of
    a valid quantization of the classical Hamiltonian, and is an important part of the full story. An unexpected property is that $\hp^a_b(x)\;g(x)^{-1/2}=0$, which would simplify matters considerably! 

As before, it may be necessary to introduce some version of regularization for these equations, but these same equations point the way to proceed. In that effort, note that although $\hp^{ac}(x)^\dag\neq \hp^{ac}(x)$ it can be helpful to know that $\hp^{ac}(x)^\dag\,\hg_{bc}(x)= \hp^{ac}(x)\,\hg_{bc}(x)$; see footnote 1.
    
    A full quantization of gravity must deal with first and likely  second order constraints, which are designed to reduce the overall Hilbert space to secure a final quantization. This paper is not the proper place to finalize a quantization of gravity. 
   
   For those who like path integration, a recent paper has found that affine quantization provides a royal path to a valid path integration of Einstein's gravity \cite{789}, in which dealing with constraints is included, along with  additional references therein.
    
    \section{Summary}
    We have stressed the definition, procedures, and advantages of affine quantization in offering to secure valid quantizations of several different examples. We first used simple models with different coordinate spaces to quantize models that fit their coordinate space in order to insure a valid result. Following that path, we examined field models by letting affine variables remove all f-issues as unphysical for any of nature's fields. Finally, we were able to put affine procedures to work on an f-issue-free version of an affine quantization of gravity, a contribution that has been needed for a long tine. It is hoped that readers can use affine quantization procedures to help solve some of their problems.

     \end{document}